\documentstyle[mprocl,epsf]{article}

\newcommand{\noi}{\noindent}

\newcommand{\beq}{\begin{eqnarray}}
\newcommand{\eeq}{\end{eqnarray}}

\newcommand{\be}{\begin{eqnarray}}
\newcommand{\en}{\end{eqnarray}}
\newcommand{\JPSJ}{J. Phys. Soc. Jpn.}

\newcommand{\PRL}{ Phys. Rev. Lett.}
\newcommand{\PRB}{ Phys. Rev. {\bf  B}}

\def\PRL{\em Phys. Rev. Lett.}

\begin{document}

\title{One-Particle vs. Two-Particle Crossover in Weakly Coupled Hubbard
Chains and Ladders:
Perturbative Renormalization Group Approach}

\author{Jun-ichiro KISHINE and Kenji YONEMITSU}

\address{Department of Theoretical Studies,
Institute for Molecular Science,\\
Okazaki 444, Japan}   


\maketitle\abstracts{Physical nature of  dimensional crossovers 
in  weakly coupled Hubbard chains and  ladders has been discussed within the framework of 
the perturbative renormalization-group approach.
The difference between these two cases originates from different universality 
classes which the corresponding isolated systems belong to.}
\noi
\vspace{-8pt}
\section{Introduction}
\vspace{-8pt}
Physical nature of dimensional crossovers in  coupled one-dimensional systems has currently
provoked a great deal of controversy.\cite{PWA,BC,VMY,BBT}
Recent discovery of the superconductivity  in the doped spin ladder under   pressure\cite{AkimitsuG} has 
stimulated us to study the dimensional crossover problem in the weakly coupled ladder system.\cite{DC,KY}
In the present work, based on the perturbative renormarization-group approach (PRG) developed by
 Bourbonnais and Caron,\cite{BC,BBT}
we discuss the dimensional crossovers induced by the inter-chain/ladder one-particle hopping, $t_{\perp}$,
 in the weakly coupled  chains or ladders, with emphasis on
 the  differences between these systems.

\section{One-Particle and Two-Particle Crossovers}
\vspace{-8pt}
The isolated chain or ladder system is specified by  the linearized bands and
scattering processes generated by the intra-chain/ladder Hubbard repulsion, $U$.
For the chain system shown in Fig.~1(a),
the scattering processes are specified by dimensionless coupling constants $g^{(1)}$ and  
$g^{(2)}$  denoting backward and forward scatterings, respectively.
For the ladder system  shown in Fig.~1(b), 
there are additional band indices [$B$ (bonding) and $A$ (anti-bonding)] and   flavor
 indices\cite{RG} ($\mu=0,f,t$), specifying the bands and scattering processes, respectively.
For the chain and ladder systems, the usual coupling constants with 
 dimension of the interaction energy are $\pi v_{F}g^{(i)}$ and $2\pi v_{F}g_{\mu}^{(i)}$, respectively,
where $v_{F}$ is the Fermi velocity at the Fermi points.

When we switch on  $U$ and  inter-chain/ladder one-particle hopping, $t_{\perp}$, as perturbations to the isolated chains/ladders, 
 dimensional crossovers are governed by the competition between the  one-particle
  and two-particle processes.\cite{BC}
To study the competition,  we set up   scaling 
equations for the inter-chain/ladder one-particle  and two-particle 
hopping amplitudes,\cite{BC,VMY,BBT} which are depicted in Figs.~2(a) and 2(b), respectively.
We parametrize the band-width cutoff as $E(l)=E_{0}e^{-l}$ with the scaling parameter, $l$.

\vspace{-16pt}
\section{Weakly Coupled Chains}
\vspace{-8pt}
The low-energy asymptotics  of the non-half filled isolated Hubbard chain
 is characterized by the {\it weak coupling} fixed point:\cite{Solyom}
$
g^{(1)\ast}=0,\,\,\,g^{(2)\ast}= U/2\pi v_{F},\label{FPC}
$
which leads the system to the {\lq\lq}Tomonaga-Luttinger liquid{\rq\rq} universality class.
The scaling equation for the interchain one-particle hopping amplitude [Fig.~2(a)] is written as
\begin{eqnarray}
{d\ln t_{\perp}(l)\over dl}=1-{1\over 4}\left[g^{(1)2}+g^{(2)2}-g^{(1)}g^{(2)}\right],
\end{eqnarray}
which gives
$
{d\ln t_{\perp}(l)/ dl}\stackrel{l\to\infty}{\longrightarrow}1-U^{2}/16\pi^2v_{F}^2. 
$
Thus  $t_{\perp}(l)$ becomes {\it relevant}    and consequently
the one-particle crossover  dominates the two-particle crossover, at least for   weak
  repulsion, $U/\pi v_{F}<1$, (see Fig.~12 of Ref.[2]) where the PRG scheme is reliable.
The one-particle crossover temperature is defined by
$
T_{\rm cross}=E_{0}e^{-l_{\rm cross}}=E_{0}[t_{\perp}/E_{0}]^{1/(1-\theta)},
$
where
$l_{\rm cross}$ is determined through 
$t_{\perp}(l_{\rm cross})=E_{0}$.\cite{BC}
The  exact solution\cite{exact} tells us the anomalous exponent, $\theta$, satisfies $\theta\leq 1/8$, which
again indicates $t_{\perp}$ is always relevant.

In Fig.~3(a), we show the  phase diagram of weakly coupled chains,
in terms of 
the intrachain Hubbard repulsion, $\tilde U=U/\pi v_{F}$, and the temperature,
$\tilde T=T/E_{0}$ for an  initial value of the interchain one-particle hopping, 
$t_{\perp}=0.01E_{0}$. A similar relust has already been given in Ref.[3]. 
The one-particle crossover dominates 
the two-particle crossover
for a wide region of  $U$. 
The  Tomonaga-Luttinger liquid (TL) phase crosses over to a two-dimensional 
 phase (2D phase) via the one-particle processes.
We show for guidence the temperature, $T_{\rm SDW}$,
 at which the amplitude of the interchain two-particle process of the SDW channel diverges.
This conclusion, however,  has excited lots of  controversy.\cite{PWA}
Anderson suggested\cite{PWA} that an intrachain repulsion of intermediate strength might 
be sufficient to cause a confinement of the particles 
within the chain. This subject, however, is beyond the scope of the PRG approach 
and here we leave  the matter open.

\section{ Weakly Coupled Ladders}
\vspace{-8pt}
Dimensional crossovers in the weakly coupled ladders\cite{KY} are very different from those in  
the weakly coupled ladders, since the  low-energy asymptotics  of  the isolated Hubbard ladder is
characterized by the {\it strong coupling} fixed point:\cite{RG}
$
g^{(1)\ast}_{0}=-1,\,\,g^{(1)\ast}_{f}=0,\,\,g^{(1)\ast}_{t}=1,
g^{(2)\ast}_{0}=-3/4+U/8\pi v_{F},\,\,g^{(2)\ast}_{f}=1/4+U/8\pi v_{F},\,\,g^{(2)\ast}_{t}=1,
$
which leads the system to the {\lq\lq}spin gap metal{\rq\rq} (SGM) phase.
The scaling equation for the interladder one-particle hopping amplitude  [Fig.~2(a)] is written as
\begin{eqnarray}
{d\ln t_{\perp}(l)\over dl}=1-\left[g_{0}^{(1)2}
    \!\!+\!\!{{ g}_{0}^{(2)2}}
    \!\!+\!\!{{ g}_{f}^{(1)2}}
    \!\!+\!\!{{ g}_{f}^{(2)2}}
    \!\!+\!\!{{ g}_{t}^{(1)2}}
   \!\!+\!\!
   {{ g}_{t}^{(2)2}}
  \!\!+\!\!{ g}_{0}^{(1)}
     { g}_{0}^{(2)}
    \!\!+\!\!{ g}_{f}^{(1)}
     { g}_{f}^{(2)}
    \!\!+\!\!{ g}_{t}^{(1)}
     { g}_{t}^{(2)}\right], 
\end{eqnarray}
which gives
$
{d\ln t_{\perp}(l)/ dl}\stackrel{l\to\infty}{\longrightarrow}-U^{2}/32\pi^2v_{F}^2-{7/8}.\label{eqn:DLcase}
$
Thus $t_{\perp}(l)$ {\it becomes  always irrelevant}.
However   {\it it can grow at an early stage of scaling } before the 
intraladder couplings  grow sufficiently. Then the competition between the one-particle and two-particle crossovers
takes place.
%%%%%%%%%%%%%%%%%%%%%%%%%%%%%%%%%%%%%%%%%%%%%%%%%%%%%%%%%%%%%%%%%%%%%%%%

In  the weakly coupled ladders, the two-particle  process is dominated by  the 
interladder   hopping (Josephson tunneling) of   $d$-wave like Cooper pairs.\cite{KY}
The lowest order   scaling equation for the   hopping amplitude, $V^{\rm SCd}$, which is depicted in 
Fig.~2(b),  is written as
\begin{eqnarray}
%%%%%%%%%%
{dV^{\rm SCd}(l)\over dl}
%&=&
=
-\left[t_{\perp}(l)g^{\rm SCd}(l)\over E_{0}\right]^{2}
%\non\\
%&+&
+
2g^{\rm SCd}(l) V^{\rm SCd} (l)
- {1\over 2}\left[V^{\rm SCd}(l)\right]^{2},  \label{eqn:scalingV}
%%%%%%%%%%
\end{eqnarray}
where $g^{{\rm SCd}}={1\over 2}(g^{(1)}_{t}+g^{(2)}_{t}-g^{(1)}_{0}-g^{(2)}_{0})$ denotes the coupling for the SCd  pair field.
We have solved (\ref{eqn:scalingV}) with the initial condition,
$
V^{\rm SCd}(0)=0.
$
The third term of the r.h.s of (\ref{eqn:scalingV}) causes divergence of  ${ V}^{{\rm 
SCd}}$ at a 
critical scaling parameter, $l_{c}$, determined  by
$
{ V}^{{\rm SCd}}(l_{c})=-\infty,
$
which gives the $d$-wave superconducting transition temperature, $T_{c}=E_{0}e^{-l_{c}}$.

In Fig.3~(b),  we show the  phase diagram of weakly coupled ladders under the same conditions
as in Fig.3(a).
In this case, the one-particle crossover is strongly suppressed due to the intraladder correlation
effects and $T_{\rm cross}$
does not exist for  larger $\tilde U$ than some crossover value, $\tilde U_{c}\sim 0.8$.
For $U<U_{c}$, the SGM phase crosses over to the two-dimensional 
 phase (2D phase) via the one-particle processes.
Then the interladder  {\it coherent} band motion takes places.
For $U>U_{c}$, the SGM  phase transits 
to the $d$-wave superconducting phase (SCd phase) via the two-particle processes.
At $T<T_{c}$, coherent   Josephson tunneling of 
the Cooper pairs in the interladder transverse direction occurs, while
 the  interladder one-particle motion is {\it incoherent}.
In both of coupled chains and ladders, in the temperature region, $T<T_{\rm cross}$, the physical properties of the 
system would strongly depend on the shape of the 2D Fermi surface.

\section{Conclusion}
\vspace{-8pt}
In the present work, we discussed nature of the dimensional crossovers in the weakly coupled chains and ladders, with emphasis on
the difference between  the two cases  within the framework of the PRG approach.
The difference of the universality class of the isolated chain and ladder profoundly affects the relevance or irrelevance of the
inter-chain/ladder one-particle hopping.
The strong coupling phase of the isolated ladder  makes the one-particle process
 {\it irrelevant} so that
the $d$-wave superconducting transition can be  induced 
via the two-particle crossover in the weakly coupled ladders.
The weak coupling phase of the isolated chain makes the one-particle process {\it relevant}
 so that the two-particle crossover
can hardly be realized in the coupled chains.

J.K was   supported by a Grant-in-Aid for Encouragement for Young Scientists from the Ministry of Education, Science, Sports and Culture, Japan. 
\vspace{-8pt}
\section*{References}
\vspace{-8pt}
%%%%%%%%%%%%%%%%%%%%%%%%%%%%%%%%%%%%%%%%%%%%%%%%%%%%%%%%%%%%%%%%%%%%%%%%%%%%%%%%%%%%%%%%%%%%%%%%%%%%

%%%%%%%%%%%%%%%%%%
\begin{figure}[h]
\caption{Linearized bands and scattering processes in the isolated
chain (a) and ladder (b).
Solid and broken lines represent the propagators of   right-moving (R) and left-moving (L) electrons, respectively.
In (b), $m$ and $\bar m$ denote different bands}
\end{figure}
%%%%%%%%%%%%%%%%%%
%%%%%%%%%%%%%%%%%%
\begin{figure}[h]
\caption{Diagrammatic representations of the scaling equations for the inter-chain/ladder ladder one-particle hopping
 amplitude (a), and 
inter-chain/ladder two-particle hopping amplitude (b). In (b),  only  he suerconducting channel is shown. 
A black circle and a shaded square represent an {\it intra}-chain/ladder scattering process and an
{\it inter}-chain/ladder two-particle hopping amplitude, respectively.
}
\end{figure}
%%%%%%%%%%%%%%%%%%
%%%%%%%%%%%%%%%%%%
\begin{figure}[h]
\caption{Phase diagram of the weakly coupled Hubbard chains (a) and ladders (b),
in terms of  the intra-chain/ladder Hubbard repulsion, $\tilde U=U/\pi v_{F}$, and the temperature,
$\tilde T=T/E_{0}$ for $t_{\perp}/E_{0}=0.001$.}
\end{figure}
%%%%%%%%%%%%%%%%%%

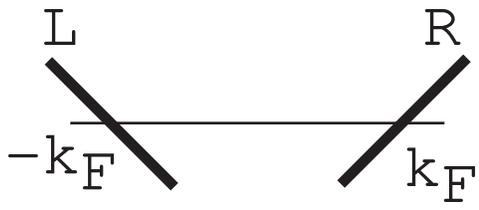
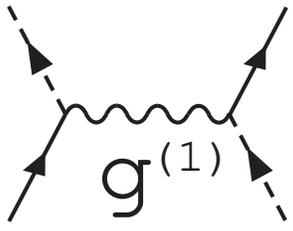
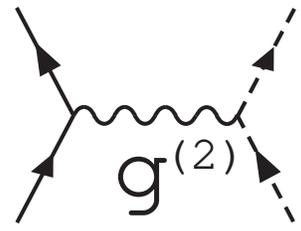
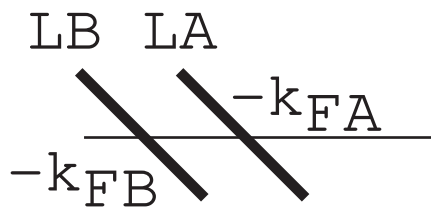
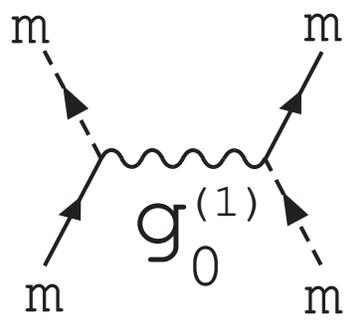
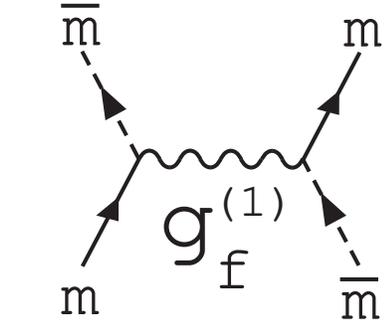
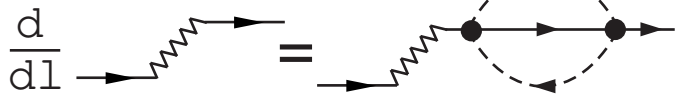
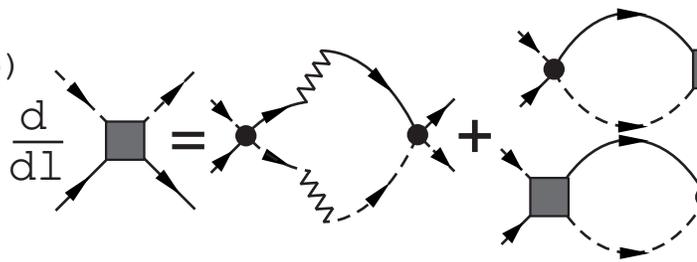
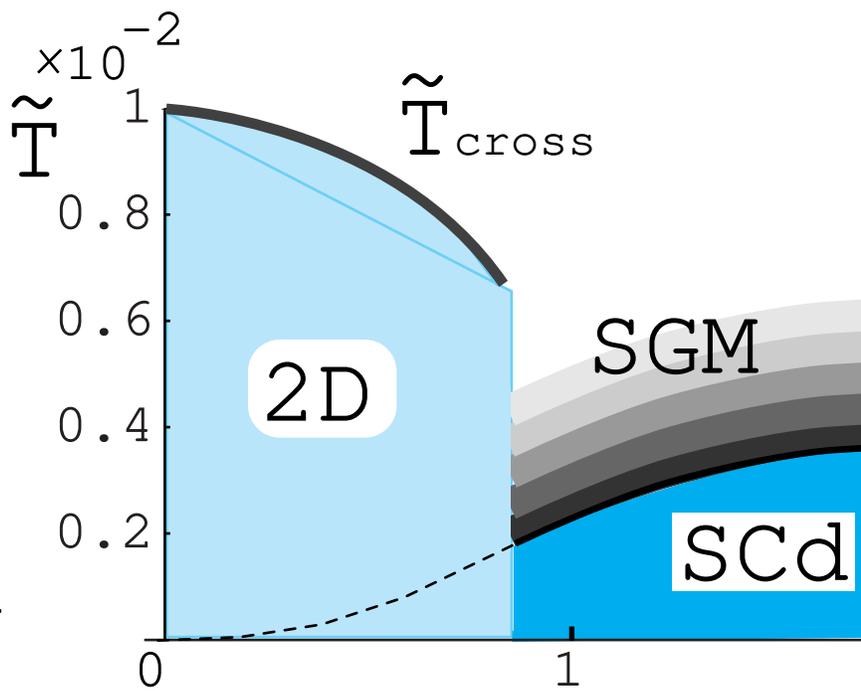
\end{document}